\documentclass[showpacs,preprintnumbers,amsmath,amssymb,aps,prd,nofootinbib,eqsecnum,tightenlines,a4paper]{revtex4}
\usepackage{graphicx,epsf}
\def\be{\begin{equation}}
\def\ee{\end{equation}}
\def\bea{\begin{eqnarray}}
\def\eea{\end{eqnarray}}
\def\p{\partial}

\def\cs2{c_{\rm{s}}^2}
\def\csa{c_{\rm{\alpha}}^2}
\def\wt{\widetilde}
\newcommand\eq[1]{Eq.~(\ref{#1})}
\newcommand\eqs[1]{Eqs.~(\ref{#1})}
\def\({\left(}
\def\){\right)}
\def\[{\left[}
\def\]{\right]}
\def\fnl{f_{\rm{NL}}}
\def\pin{p_{\rm{in}}}
\def\Omdec{\Omega_{\sigma\rm{dec}}}
\def\zetag{{\zeta_{\rm{g}}}}
\def\zetasb{\zeta_{2\rm{SB}}}
\def\mpl{M_{\rm{pl}}^2}
\def\se{\sigma_{\rm{e}}}
\begin{document}
%\preprint{}
\title{A numerical study of non-gaussianity in the curvaton scenario}
\author{Karim A.~Malik and David H.~Lyth} 
\affiliation{Cosmology and Astroparticle Physics Group, Department of
Physics, University of Lancaster, Lancaster LA1 4YB, United Kingdom }
\date{\today}
\begin{abstract}
We study the curvaton scenario using gauge-invariant second order
perturbation theory and solving the governing equations
numerically. Focusing on large scales we calculate the non-linearity
parameter $\fnl$ in the two-fluid curvaton model and compare our
results with previous analytical studies employing the sudden decay
approximation. We find good agreement of the two approaches for large
curvaton energy densities at curvaton decay, $\Omdec$, but significant
differences of up to $10\%$ for small $\Omdec$.
\end{abstract}

\pacs{98.80.Cq \hfill JCAP09 (2006) 008, astro-ph/0604387v3}

\maketitle

%%%%%%%%%%%%%%%%%%%%%%%%%%
\section{Introduction}
%%%%%%%%%%%%%%%%%%%%%%%%%%

The third year WMAP data release \cite{WMAP} has confirmed beautifully
the cosmological standard model of structure formation: during
inflation fluctuations in the scalar fields are stretched to
super-horizon scales and later on source the Cosmic Microwave Background
(CMB) anisotropies and the large scale structure.
In the standard inflation models the scalar field responsible for the
accelerated expansion of the universe, the inflaton, also provides
these fluctuations \cite{LLBook}.
Recently a related but different scenario has become become popular:
the curvaton paradigm
\cite{curvaton,MT,Enqvist,LUW,MWU,Lyth:2003dt,GMW,Enqvist:2005pg,
Linde2005,Lyth2006}. Here the fluctuations are not generated by the
inflaton, but by a different scalar field, the curvaton.

A powerful tool to differentiate between different models of the early
universe is second order perturbation theory
\cite{Mukhanov,Bruni,Maldacena,Acquaviva,Nakamura,Noh,Bartolo:2001cw,
Bernardeau:2002jy,Bernardeau:2002jf,MW2004,Bartolo2003,Bartolo:2004if,
Bartolo:2004ty,Enqvist:2004bk,RiSh,Rigopoulos:2004ba,Bartolo:JCAP,
Tomita:2005et,Lyth:2005du,Seery:2005gb,M2005}. It allows the accurate
calculation of higher order statistics such as the primordial
bispectrum and the non-linearity parameter $\fnl$.
Instead of using cosmological perturbation theory at second order, the
$\Delta N$-formalism \cite{SB,SaSt95,SaTa,LMS,Lyth:2005fi,LaVe} has
proved useful to study non-gaussianity and calculate $\fnl$.

Bartolo et al.~\cite{Bartolo2003} studied the curvaton model at second
order using cosmological perturbation theory, and Lyth and Rodriguez
used the $\Delta N$-formalism to calculate the non-linearity parameter
$\fnl$ in this scenario. However, both studies used the sudden decay
approximation. Here we go beyond sudden decay, using second order
gauge-invariant perturbation theory and solve the ensuing equations
numerically.

We consider scalar perturbations up to and including second order and
assume a flat Friedmann-Robertson-Walker (FRW) background
spacetime. We work on large scales (compared to the horizon size),
which allows us to neglect gradient terms.

The outline of the paper is as follows. In the next section we give
the governing equations up to second order and define the
gauge-invariant variables we are using. We specify in Section
\ref{model_sect} the two-fluid curvaton model we are studying and
apply the equations of Section \ref{gov_sect}. In Section
\ref{deltaN_sect} we take a small detour from perturbation theory and
compare our perturbative approach to the $\Delta N$ formalism.
After defining the non-linearity parameter $\fnl$ we present numerical
solutions in Section \ref{result_sect} and compare our numerical
results to the sudden decay approximation.
The governing equations without any gauge restrictions are given in
the appendix.

%%%%%%%%%%%%%%%%%%%%%%%%%%%%%%%%%%%%%%%%%%%%%
\section{Governing equations}
\label{gov_sect}
%%%%%%%%%%%%%%%%%%%%%%%%%%%%%%%%%%%%%%%%%%%%%

In this section we give the governing equations for a system of
multiple interacting fluids on large scales, allowing for scalar
perturbations up to second order, following closely the treatment of
Refs.~\cite{MW2004} and \cite{MW2005}.

The covariant Einstein equations are given by\footnote{Notation: Greek
indices, $\mu,\nu,\lambda$, run from $0,\ldots3$, while lower case
Latin indices, $i,j,k$, run from $1,\ldots3$. Greek indices from the
beginning of the alphabet, $\alpha,\beta,\gamma$ will be used to
denote different fluids.}
\be
\label{Einstein}
G_{\mu\nu}=8\pi G \; T_{\mu\nu} \,,
\ee
where $G_{\mu\nu}$ is the Einstein tensor, $T_{\mu\nu}$ is the
total energy-momentum tensor, and $G$ is Newton's constant.
Through the Bianchi identities, the field equations (\ref{Einstein})
imply the local conservation of the total energy and momentum,
\be
\label{nablaTmunu}
\nabla_\mu T^{\mu\nu}=0\,.
\ee

In the multiple fluid case the total energy-momentum tensor
is the sum of the energy-momentum tensors of the 
individual fluids
\be
T^{\mu\nu}=\sum_\alpha T^{\mu\nu}_{(\alpha)}\,.
\ee
For each fluid we define the local energy-momentum transfer 4-vector
$Q^\nu_{(\alpha)}$ through the relation
\be
\label{nablaTalpha}
\nabla_\mu T^{\mu\nu}_{(\alpha)}=Q^\nu_{(\alpha)}\,,
\ee
where energy-momentum is locally conserved for $Q^\nu_{(\alpha)}=0$,
i.e.~only for non-interacting fluids. Equations~(\ref{nablaTmunu})
and~(\ref{nablaTalpha}) imply the constraint
\be
\label{Qconstraint}
\sum_\alpha Q^\nu_{(\alpha)}=0 \,.
\ee

We split scalar perturbations into background, first, and second order
quantities according to, 
\be
\label{rhosplit}
\rho(x^\mu)\equiv\rho_0(t)+\delta\rho_1(x^\mu)
+\frac{1}{2}\delta\rho_2(x^\mu)\,,
\ee
using here the total energy density as an example.

The line element on large scales is given by
\be 
\label{ds2}
ds^2=-\[1+2\(\phi_1+\frac{1}{2}\phi_2\)\]dt^2
+a^2\[1-2\(\psi_1-\frac{1}{2}\psi_2\)\]\delta_{ij}dx^idx^j \,, 
\ee
where $a=a(t)$ is the scale factor, $\phi_1$ and $\phi_2$ are the
lapse functions at first and second order, respectively, and $\psi_1$
and $\psi_2$ the curvature perturbations.

Following Refs.~\cite{KS,MW2005} we split the energy-momentum transfer
4-vector using the total fluid velocity $u^\mu$ as
\be
Q^\mu_{(\alpha)}\equiv Q_{(\alpha)} u^\mu+f^\mu_{(\alpha)}\,,
\ee
where $Q_{(\alpha)}$ is the energy transfer rate and
$f^\mu_{(\alpha)}$ the momentum transfer rate, subject to the
condition
$
u_\mu f^\mu_{(\alpha)}=0
$.

On large scales the only non-zero component of the 4-velocity is
\bea
u_0 &=& -\left[1+\phi_1+\frac{1}{2}\phi_2-\frac{1}{2}\phi_1^2
\right]\,, \\
\eea
subject to the constraint
$u_\mu u^\mu=-1$.
We then find the only non-zero component of the energy transfer
4-vector on large scales to be
\be
Q_{0(\alpha)}=
-Q_{0\alpha}\(1+\phi_1+\frac{1}{2}\phi_2-\frac{1}{2}\phi_1^2\)
-\delta Q_{1\alpha}\(1+\phi_1\)-\delta Q_{2\alpha}\,,
\ee
where $Q_{0\alpha}$, $\delta Q_{1\alpha}$, and $\delta Q_{2\alpha}$
are the energy transfer to the $\alpha$-fluid in the background, at
first and at second order, respectively.

%%%%%%%%%%%%%%%%%%%%%%%%%%
\subsection{Background}
%%%%%%%%%%%%%%%%%%%%%%%%%%

Energy conservation for the $\alpha$-fluid in the background is given
from \eq{nablaTalpha} as
\be
\label{dotrho0alpha}
\dot\rho_{0\alpha}=-3H\(\rho_{0\alpha}+P_{0\alpha}\)+Q_{0\alpha}\,,
\ee
where $H=\dot a/a$ is the Hubble parameter, and $\rho_{0\alpha}$ and
$P_{0\alpha}$ are the energy density and the pressure, respectively,
of the $\alpha$-fluid.
Total energy conservation is then given by summing over the individual
fluids and using \eq{Qconstraint} as
\be
\dot\rho_{0}=-3H\(\rho_{0}+P_{0}\)\,,
\ee
where $\rho_0=\sum_\alpha \rho_{0\alpha}$.

The Friedmann constraint is given from the $0-0$ component \eq{Einstein} as
\be
\label{Friedmann}
H^2 = \frac{8\pi G}{3}\rho_0 \,.
\ee
%

%%%%%%%%%%%%%%%%%%%%%%%
%\section{Perturbations}
%%%%%%%%%%%%%%%%%%%%%%%

%%%%%%%%%%%%%%%%%%%%%%%%%%%%%%%%%%%%%%
\subsection{First order perturbations}
%%%%%%%%%%%%%%%%%%%%%%%%%%%%%%%%%%%%%%

We now give the governing equations on large scales at first order in
the perturbations in the flat gauge, denoting quantities evaluated in
this gauge by a ``tilde''. The governing equations in an arbitrary
gauge are given in appendix \ref{governing1}.

The energy conservation equation for the $\alpha$-fluid at first order is
given from \eq{nablaTalpha} on large scales as
\be 
\label{dotrho1alpha}
\dot{\wt{\delta\rho_{1\alpha}}}
+3H\left( \wt{\delta\rho_{1\alpha}}+\wt{\delta P_{1\alpha}}\right)
-Q_{0\alpha}\wt{\phi_1} - \wt{\delta Q_{1\alpha}}=0\,.
\ee
The total energy density perturbation is related to the individual fluid
densities, and similarly for the pressure perturbations, by
\bea 
\label{defrhotot}
\wt{\delta\rho_1}&\equiv&\sum_\alpha \wt{\delta\rho_{1\alpha}}\,,\qquad
\wt{\delta P_1}\equiv\sum_\alpha \wt{\delta P_{1\alpha}}\,,
\eea
and we get, using the constraint \eq{Qconstraint}, from
\eq{dotrho1alpha} the evolution equation for the total energy density
perturbation
\be 
\dot{\wt{\delta\rho_{1}}}
+3H\left( \wt{\delta\rho_{1}}+\wt{\delta P_{1}}\right)
=0\,.
\ee

The $0-0$ Einstein equation on flat slices is, using \eq{Einstein} and
the background Friedmann constraint (\ref{Friedmann}), given by
\be
\wt\phi_1
= -\frac{1}{2}\frac{\wt{\delta\rho_1}}{\rho_0}\,.
\ee

The curvature perturbation on uniform $\alpha$ fluid energy density
hyper-surfaces at first order is given by \cite{WMLL}
\be
\label{zeta1alpha}
\zeta_{1\alpha}=
-H\frac{\delta\rho_{1\alpha}}{\dot\rho_{0\alpha}}\,.
\ee
The curvature perturbation on uniform total energy density
hyper-surfaces at first order is given by
\be
\label{zeta1}
\zeta_1=-\frac{H}{\dot\rho_0}\wt{\delta\rho_1}\,,
\ee
and related to the curvature perturbation on uniform $\alpha$ fluid
slices by
\be
\label{zeta1relate}
\zeta_1=
\sum_\alpha\frac{\dot\rho_{0\alpha}}{\dot\rho_{0}}
\zeta_{1\alpha}\,.
\ee

As in the background we introduce new variables, the normalised
energy densities at first order, 
\be
\label{defdeltaOmega1}
\delta\Omega_{1\alpha}\equiv\frac{\wt{\delta\rho_{1\alpha}}}{\rho_0}\,,
\ee
which allow us in combination with choosing the number of e-foldings
as a time variable to write the governing equations in the following
sections in a particularly compact form.

In terms of the new variables the curvature perturbation on uniform
$\alpha$ fluid energy density hyper-surfaces, given in
\eq{zeta1alpha}, is simply
\be
\zeta_{1\alpha}=
-H\frac{\rho_0}{\dot\rho_{0\alpha}}\delta\Omega_{1\alpha}\,.
\ee

%%%%%%%%%%%%%%%%%%%%%%%%%%
\subsection{Second order}
%%%%%%%%%%%%%%%%%%%%%%%%%%

We now give the governing equations on large scales at second order in
the perturbations in the flat gauge, denoting quantities evaluated in
this gauge by a ``tilde''.  The governing equations in an arbitrary
gauge are given in appendix \ref{governing2}.

The energy conservation equation for the $\alpha$-fluid at second
order is given from \eq{nablaTalpha} on large scales by
\be 
\label{dotrho2alpha}
\dot{\wt{\delta\rho_{2\alpha}}}
+3H\left( \wt{\delta\rho_{2\alpha}}+\wt{\delta P_{2\alpha}}\right)
-Q_{0\alpha}\left(\wt{\phi_2}-{\wt{\phi_1}}^2\right) - \wt{\delta Q_{2\alpha}}
-2\phi_1\wt{\delta Q_{1\alpha}}
=0\,.
\ee
Using \eq{Qconstraint} the evolution equation for the total energy
density is
\be 
\dot{\wt{\delta\rho_{2}}}
+3H\left( \wt{\delta\rho_{2}}+\wt{\delta P_{2}}\right)
=0\,,
\ee
where the total density and pressure perturbations are given in terms of the
individual fluid ones by
\bea
%\wt{\delta\rho_1}&\equiv&\sum_\alpha \wt{\delta\rho_{1\alpha}}\\
\wt{\delta\rho_2} \equiv \sum_\alpha \wt{\delta\rho_{2\alpha}}\,,\qquad
\wt{\delta P_2} \equiv \sum_\alpha \wt{\delta P_{2\alpha}}\,.
\eea

The $0-0$ Einstein equation on flat slices is, using \eq{Einstein} and
the background Friedmann constraint (\ref{Friedmann}), given by
\be
\wt\phi_2-4\wt{\phi_1}^2
= -\frac{1}{2}\frac{\wt{\delta\rho_2}}{\rho_0}\,.
\ee

The curvature perturbation at second order in terms of uniform
$\alpha$-density perturbations on flat slices is given by
\cite{MW2004,M2005}
\be
\label{defzeta2alpha}
\zeta_{2\alpha}=-\frac{H}{\dot\rho_{0\alpha}}\wt{\delta\rho_{2\alpha}}
+2\frac{H}{{\dot\rho_{0\alpha}}^2}\dot{\wt{\delta\rho_{1\alpha}}}
\wt{\delta\rho_{1\alpha}}
+\frac{H}{{\dot\rho_{0\alpha}}^2}
\left[
H\(5+3\csa\)+\frac{\dot H}{H}\frac{Q_{0\alpha}}{\dot\rho_{0\alpha}}
-\frac{\dot Q_{0\alpha}}{\dot\rho_{0\alpha}}
\right]{\wt{\delta\rho_{1\alpha}}}^2\,,
\ee
where $\csa\equiv\dot P_{0\alpha}/\dot\rho_{0\alpha}$ is the
adiabatic sound speed of the $\alpha$-fluid.

The curvature perturbation at second order in terms of the total density
perturbations on flat slices is given by \cite{MW2004}
\be
\label{defzeta2}
\zeta_2=-\frac{H}{\dot\rho_0}\wt{\delta\rho_2}
+2\frac{H}{{\dot\rho_0}^2}\dot{\wt{\delta\rho_1}}\wt{\delta\rho_1}
+\frac{H^2}{{\dot\rho_0}^2}\left(5+3\cs2\right){\wt{\delta\rho_1}}^2\,,
\ee
where $\cs2\equiv\dot P_{0}/\dot\rho_{0}$ is the total adiabatic sound
speed related to the individual speeds $\csa$ by
\be
\label{cs2}
\cs2=\sum_\alpha\frac{\dot\rho_{0\alpha}}{\dot\rho_0}\csa\,.
\ee

As at first order we introduce new variables, the normalised energy
densities at second order, allowing us to rewrite the governing
equations in the following sections in a particularly compact form,
\be
\label{defdeltaOmega2}
\delta\Omega_{2\alpha}\equiv\frac{\wt{\delta\rho_{2\alpha}}}{\rho_0}\,.
\ee

The curvature perturbation $\zeta_2$, defined above in \eq{defzeta2},
is related to the curvature perturbation employed in the $\Delta N$
formalism (see also \cite{LW}), which we denote by $\zetasb$, by
\cite{LMS,Lyth:2005du}
\be
\label{defzetasb}
\zetasb=\zeta_2-2\zeta_1^2\,.
\ee
It was originally introduced by Salopek and Bond \cite{SB} and
employed by Maldacena in studies of non-gaussianity in
Ref.~\cite{Maldacena}.

%%%%%%%%%%%%%%%%%%%%%%%
\section{The model}
\label{model_sect}
%%%%%%%%%%%%%%%%%%%%%%%

In this section we specify the curvaton model and apply the governing
equations given in the previous section order by order.
%
%NEW V3 added ref.
We model the curvaton as a pressureless fluid \cite{curvaton}, and
hence our system will be governed by the equations of state
\be
\label{defEOS}
P_\sigma=0 \,, \qquad P_\gamma=\frac{1}{3}\rho_\gamma \,,
\ee
where the subscripts ``$\sigma$'' and ``$\gamma$'' denote the curvaton
and the radiation fluid, respectively.
The decay of the curvaton is described by a fixed decay rate, $\Gamma=const$, 
\be
\label{defQmodel}
Q_\sigma=-\Gamma\rho_\sigma \,, \qquad Q_\gamma=\Gamma\rho_\sigma\,,
\ee
where we used \eq{Qconstraint}.

%%%%%%%%%%%%%%%%%%%%%%%
\subsection{Background}
%%%%%%%%%%%%%%%%%%%%%%%

%
%%%%%%%%%%%%%%%%%%%%%%%%%%%
\begin{figure}
\begin{center}
\includegraphics[angle=0, width=88mm]{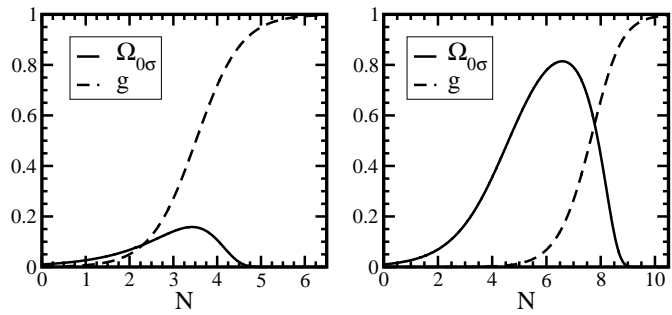} \\
\vspace{0.2cm}
\caption[pix]{\label{pix1} Evolution of the normalised background
curvaton density, $\Omega_{0\sigma}$, and the normalised decay rate
$g$, as a function of the number of e-foldings, starting with initial
density and decay rate $\Omega_{0\sigma}=10^{-2}$ and
$\Gamma/H=10^{-3}$, corresponding to $\pin=0.32$ on the left panel,
and on the right panel with $\Gamma/H=10^{-6}$, corresponding to
$\pin=10$.}
\end{center}
\end{figure}
%%%%%%%%%%%%%%%%%%%%%%%%%%
%
The background evolution equations are from \eq{dotrho0alpha} and
using \eqs{defEOS} and (\ref{defQmodel}) given by
\bea
\dot\rho_\sigma &=& 
-3H\rho_{0\sigma}-\Gamma\rho_{0\sigma}\,, \\
\dot\rho_{0\gamma} &=& -4H\rho_{0\gamma}+\Gamma\rho_{0\sigma}\,.
\eea

We now change to a new set of variables. First we introduce
normalised energy densities in the background,
\be
\Omega_{0\alpha}\equiv\frac{\rho_{0\alpha}}{\rho_0}\,,
\ee
and define the reduced decay rate as
\be
g\equiv\frac{\Gamma}{\Gamma+H}\,.
\ee
We change the time coordinate from coordinate time $t$ to the number
of e-foldings $N\equiv\ln a$, that is $\frac{d}{dt}=H\frac{d}{dN}$.
The normalised radiation energy density is then simply given from the
Friedmann equation, (\ref{Friedmann}), as 
$\Omega_{0\gamma}\equiv 1-\Omega_{0\sigma}$
and we get the system of background evolution equations in terms of
these new variables
\bea
\label{Omegadash}
\Omega_{0\sigma}'&=&\Omega_{0\sigma}
\left(1-\Omega_{0\sigma}-\frac{g}{1-g}\right)\,, \\
\label{gdash}
g'&=& \frac{1}{2}\left(4-\Omega_{0\sigma}\right)\left(1-g\right)g\,.
\eea

Solutions for the system (\ref{Omegadash}) and (\ref{gdash}) are given
in Fig.~\ref{pix1} for two different initial conditions,
$\Omega_{0\sigma}=10^{-2}$ and $\Gamma/H=10^{-3}$ and
$\Omega_{0\sigma}=10^{-2}$ and $\Gamma/H=10^{-6}$.

It was shown in Ref.~\cite{MWU} that the solutions of the system
(\ref{Omegadash}) and (\ref{gdash}) depend only on a single parameter
since we can write $\Omega_{0\sigma}=\Omega_{0\sigma}(g)$, and for
$g\ll 1$ we can solve the system explicitly, which gives
$\Omega_{0\sigma}\propto\sqrt{g}$.
We therefore define the parameter \cite{MWU,GMW}
\be
\label{pin} 
\pin\equiv\frac{\Omega_{0\sigma}}{\sqrt{g}}\Big|_{\rm{in}}\,,  
\ee
the subscript ``in'' denoting the initial conditions.
For the initial conditions $\Omega_{0\sigma}=10^{-2}$ and
$\Gamma/H=10^{-3}$ and $\Omega_{0\sigma}=10^{-2}$ and
$\Gamma/H=10^{-6}$ the parameter $\pin$ takes the values $0.32$ and
$10$, respectively.

%%%%%%%%%%%%%%%%%%%%%%%%%%
\subsection{First order}
\label{gov_equ1}
%%%%%%%%%%%%%%%%%%%%%%%%%%

%%%%%%%%%%%%%%%%%%%%%%%%%%%
\begin{figure}
\begin{center}
\includegraphics[angle=-90, width=70mm]{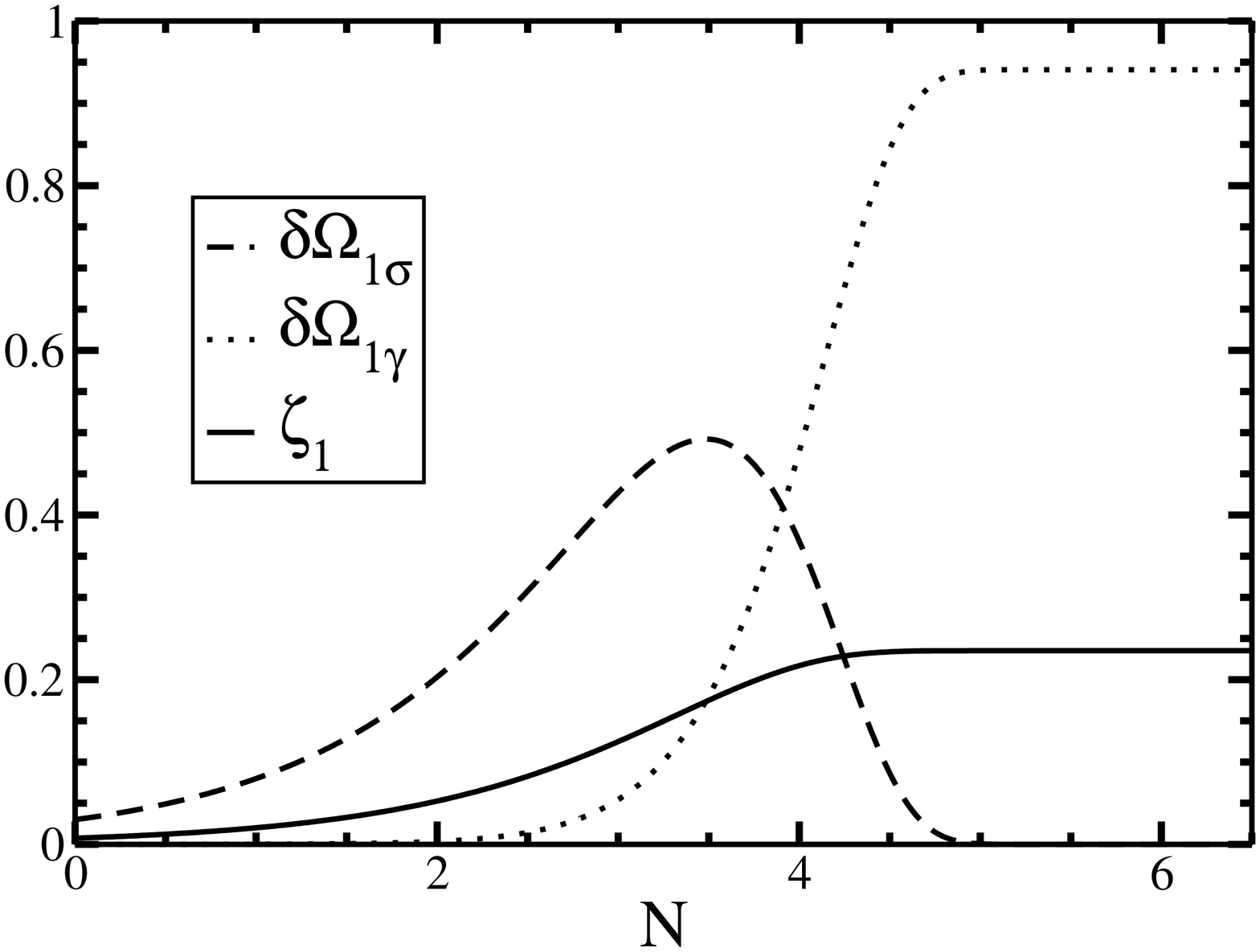} \\
\vspace{0.2cm}
\caption[pix]{
\label{pix3} 
Evolution of the total curvature perturbation, $\zeta_1$, and the
normalised density perturbations at first order as a function of the
number of e-foldings, starting with $\zeta_{1\sigma}=1$ and initial
density and decay rate $\Omega_{0\sigma}=10^{-2}$ and $\Gamma/H=10^{-3}$,
corresponding to $\pin=0.32$.}
\end{center}
\end{figure}
%%%%%%%%%%%%%%%%%%%%%%%%%%

%%%%%%%%%%%%%%%%%%%%%%%%%%%
\begin{figure}
\begin{center}
\includegraphics[angle=-90, width=70mm]{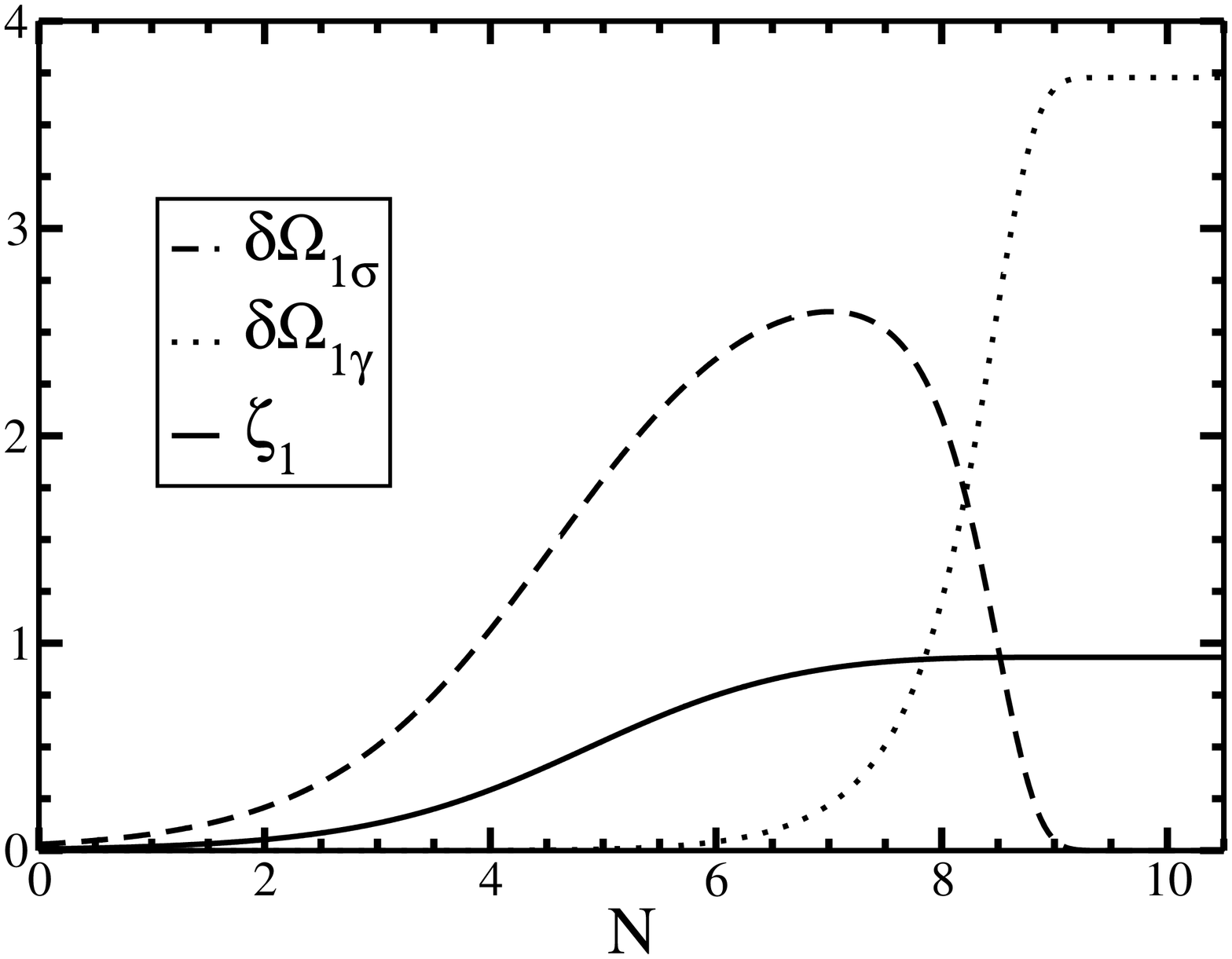} \\
\vspace{0.2cm}
\caption[pix]{
\label{pix4} 
Same as Fig.~\ref{pix3} but with $\Gamma/H=10^{-6}$ initially,
corresponding to $\pin=10$.}
\end{center}
\end{figure}
%%%%%%%%%%%%%%%%%%%%%%%%%%

The perturbed energy transfer rates are given from \eq{defQmodel} at first
order as
\be
\label{deltaQ1model}
\delta Q_{1 \sigma}=-\Gamma \delta\rho_{1\sigma}\,,
\qquad 
\delta Q_{1 \gamma}=\Gamma \delta\rho_{1\sigma}\,.
\ee
The evolution equations at first order are from \eq{dotrho1alpha} and
\eqs{defEOS}, and using (\ref{deltaQ1model}) in terms of the
normalised energy densities defined in \eq{defdeltaOmega1} given by
\bea
\label{deltaOmega1sigma}
&&\delta\Omega_{1\sigma}'+\left[
\frac{3-2g}{1-g}-\frac{\Omega_{0\sigma}}{2}\frac{6-5g}{1-g}
-4\Omega_{0\gamma}\right]\delta\Omega_{1\sigma}
-\frac{\Omega_{0\sigma}}{2}\frac{g}{1-g}\delta\Omega_{1\gamma}
=0\,,\\
\label{deltaOmega1gamma}
&&\delta\Omega_{1\gamma}'+\[4\(1-\Omega_{0\gamma}\)
-\frac{6-7g}{1-g}\frac{\Omega_{0\sigma}}{2}\] \delta\Omega_{1\gamma}
-\frac{g}{1-g}\(1- \frac{\Omega_{0\sigma}}{2} \)
\delta\Omega_{1\sigma}
=0\,.
\eea

The curvature perturbations at first order on uniform curvaton and
radiation density hypersurfaces are from \eq{zeta1alpha} given by
\bea
\label{zeta1sigma}
\zeta_{1\sigma}&=&
\frac{1-g}{(3-2g)\Omega_{0\sigma}}\delta\Omega_{1\sigma}\,,\\
\zeta_{1\gamma}&=& \frac{1-g}{4\(1-g\)\Omega_{0\gamma}-g\Omega_{0\sigma}}
\delta\Omega_{1\gamma}\,.
\eea
The curvature perturbation on uniform total density slices in terms of
the new variables is given by
\be
\label{zeta1model}
\zeta_1=\frac{\delta\Omega_{1\sigma}+\delta\Omega_{1\gamma}}
{3\Omega_{0\sigma}+4\Omega_{0\gamma}}\,.
\ee

Solutions for the equation system (\ref{deltaOmega1sigma}) and
(\ref{deltaOmega1gamma}) are given in Figs.~\ref{pix3} and \ref{pix4}
for two different sets of initial conditions,
$\Omega_{0\sigma}=10^{-2}$ and $\Gamma/H=10^{-3}$ and
$\Omega_{0\sigma}=10^{-2}$ and $\Gamma/H=10^{-6}$, corresponding to
$\pin=0.32$ and $\pin=10$, respectively. Using \eq{zeta1model} we also
plot the evolution of $\zeta_1$.
For the perturbations we use the initial conditions
\be
\label{IC_1}
\zeta_{1\sigma,\rm{in}}=1\,, \qquad \zeta_{1\gamma,\rm{in}}=0\,,
\ee
which can be easily translated in initial conditions for
$\delta\Omega_{1\sigma}$ using \eq{zeta1sigma}, and facilitates
comparison with Ref.~\cite{MWU}.

Note that the values for $\delta\Omega_{1\sigma}$ and
$\delta\Omega_{1\gamma}$ can exceed $1$, as can be seen in
Fig.~\ref{pix4}. 
%
%NEW V3
%
This doesn't indicate the ``breakdown of perturbation
theory'' or anything dramatic like it, but is merely an artifact of
normalising the density perturbations by the total background density
$\rho_0$, which can itself be small. As in the background, the
normalised energy densities together with the choice of time
coordinate give a particularly neat system of governing equations.

%%%%%%%%%%%%%%%%%%%%%%%%%%
\subsection{Second order}
\label{gov_equ2}
%%%%%%%%%%%%%%%%%%%%%%%%%%

%%%%%%%%%%%%%%%%%%%%%%%%%%
\begin{figure}
\begin{center}
\includegraphics[angle=-90, width=70mm]{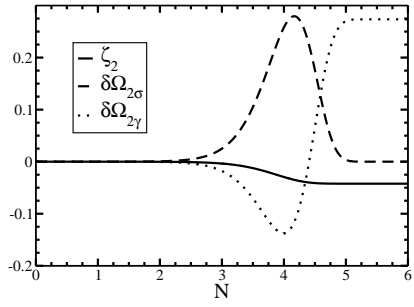} \\
\vspace{0.2cm}
\caption[pix]{
\label{pix5}
Evolution of the total curvature perturbation, $\zeta_2$, and the
normalised density perturbations at second order as a function of the
number of e-foldings, with initially $\zeta_{1\sigma,\rm{in}}=1$ and
density and decay rate $\Omega_{0\sigma}=10^{-2}$ and
$\Gamma/H=10^{-3}$, corresponding to $\pin=0.32$.}
\end{center}
\end{figure}
%%%%%%%%%%%%%%%%%%%%%%%%%%

%%%%%%%%%%%%%%%%%%%%%%%%%%%
\begin{figure}
\begin{center}
\includegraphics[angle=-90, width=70mm]{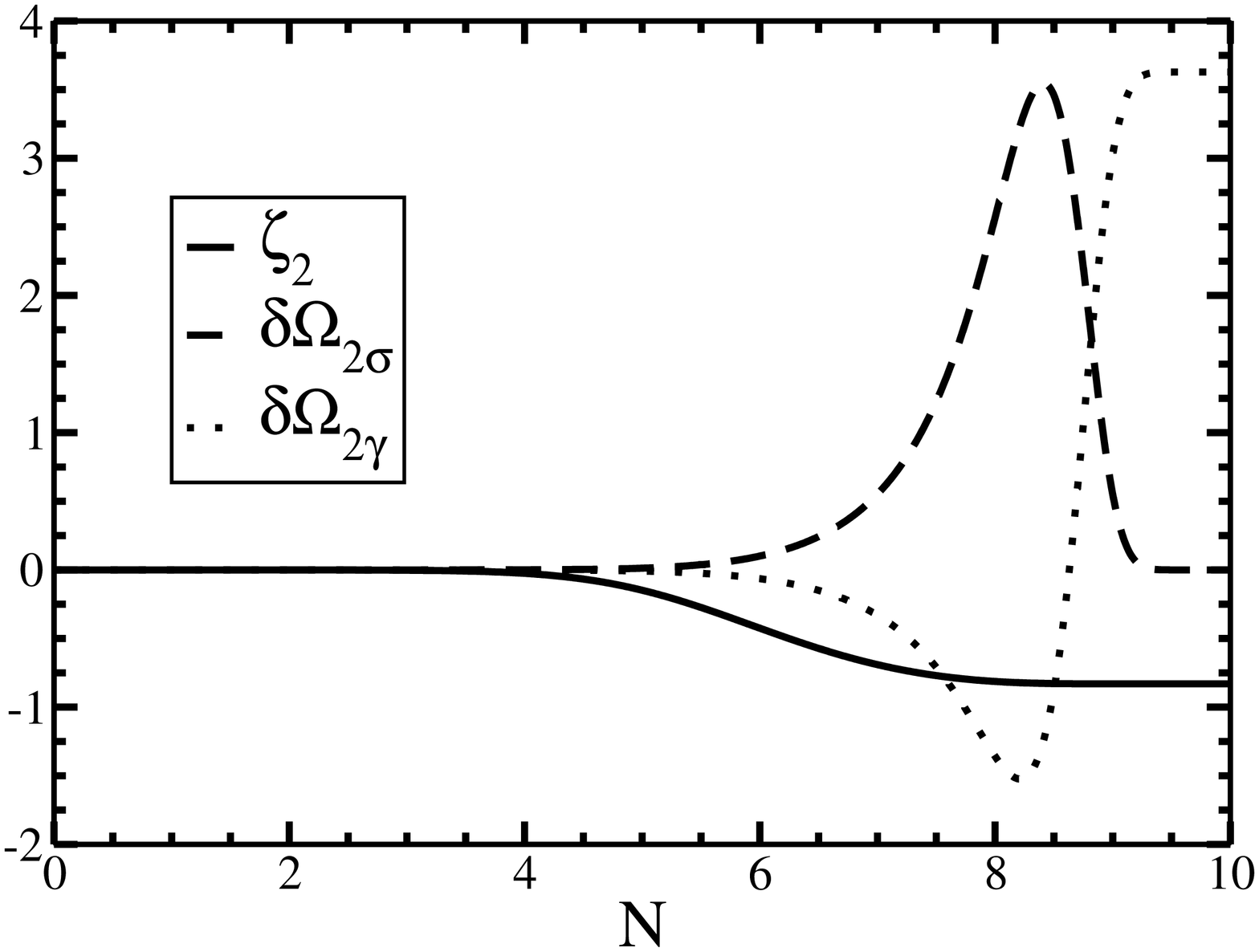} \\
\vspace{0.2cm}
\caption[pix]{
\label{pix6} 
Same as Fig.~\ref{pix5} but with $\Gamma/H=10^{-6}$ initially,
corresponding to $\pin=10$.}
\end{center}
\end{figure}
%%%%%%%%%%%%%%%%%%%%%%%%%%

The perturbed energy transfer rates are given from \eq{defQmodel} at second
order as
\be
\label{deltaQ2model}
\delta Q_{2 \sigma}=-\Gamma \delta\rho_{2\sigma}\,,
\qquad 
\delta Q_{2 \gamma}=\Gamma \delta\rho_{2\sigma}\,.
\ee
We then find evolution equations at second order from
\eq{dotrho2alpha} and using \eqs{defEOS} and (\ref{deltaQ2model}) in
terms of the normalised energy densities defined above in
\eqs{defdeltaOmega1} and (\ref{defdeltaOmega2}) to be
\bea
\label{deltaOmega2sigma}
\delta\Omega_{2\sigma}'&+&\left[
\frac{3-2g}{1-g}-\frac{\Omega_{0\sigma}}{2}\frac{6-5g}{1-g}
-4\Omega_{0\gamma}\right]\delta\Omega_{2\sigma}
-\frac{\Omega_{0\sigma}}{2}\frac{g}{1-g}\delta\Omega_{2\gamma}\nonumber\\
&&\qquad-\frac{g}{1-g}\(\delta\Omega_{1\sigma}+\delta\Omega_{1\gamma}\)\[
\delta\Omega_{1\sigma}-\frac{3}{4}\Omega_{0\sigma}
\(\delta\Omega_{1\sigma}+ \delta\Omega_{1\gamma}\)
\]
=0\,,\\
\label{deltaOmega2gamma}
\delta\Omega_{2\gamma}'&+&
\[4\(1-\Omega_{0\gamma}\)
-\frac{6-7g}{1-g}\frac{\Omega_{0\sigma}}{2}\]
\delta\Omega_{2\gamma}
-\(1-\frac{\Omega_{0\sigma}}{2}\)\frac{g}{1-g}\delta\Omega_{2\sigma}
\nonumber\\
&&\qquad+\frac{g}{1-g}\(\delta\Omega_{1\sigma}+\delta\Omega_{1\gamma}\)\[
\delta\Omega_{1\sigma}-\frac{3}{4}\Omega_{0\sigma}
\(\delta\Omega_{1\sigma}+ \delta\Omega_{1\gamma}\)
\]
=0\,.
\eea

The adiabatic sound speed in a multi-fluid system is given above in
\eq{cs2}
and we find for the two-fluid curvaton model 
\be
\label{cs2_app}
\cs2=\frac{1}{3}\frac{4\(1-g\)\Omega_{0\gamma}-g\Omega_{0\sigma}}
{\(1-g\)\(3\Omega_{0\sigma}+4\Omega_{0\gamma} \)}\,.
\ee

The curvature perturbation on uniform total density hypersurfaces at
second order in terms of the normalised quantities is
\bea
\label{zeta2model}
\zeta_2=
\frac{\delta\Omega_{2\sigma}+\delta\Omega_{2\gamma}}
{3\Omega_{0\sigma}+4\Omega_{0\gamma}}
-\(1-3\cs2\)\(\frac{
\delta\Omega_{1\sigma}+\delta\Omega_{1\gamma}
}{
3\Omega_{0\sigma}+4\Omega_{0\gamma}
}\)^2
-2\frac{\(\delta\Omega_{1\sigma}+\delta\Omega_{1\gamma}\)
\delta\Omega_{1\gamma}
}{\(3\Omega_{0\sigma}+4\Omega_{0\gamma}\)^2
}\,,
\eea
where $\cs2$ is given above in \eq{cs2_app}.

%NEW V3
The system of equations (\ref{deltaOmega2sigma}) and
(\ref{deltaOmega2gamma}) is readily integrated using a standard fourth
order Runge-Kutta solver \cite{NR}.
We give the solutions for this system of equations for the two
different sets of initial conditions, $\Omega_{0\sigma}=10^{-2}$ and
$\Gamma/H=10^{-3}$ and $\Omega_{0\sigma}=10^{-2}$ and
$\Gamma/H=10^{-6}$, corresponding to $\pin=0.32$ and $\pin=10$, in
Figs.~\ref{pix5} and \ref{pix6}. Using \eq{zeta2model} we also plot
the evolution of $\zeta_2$.
The initial conditions for the second order perturbations are chosen as
\be
\label{IC_2}
\delta\Omega_{2\sigma,\rm{in}}=0\,, \qquad 
\delta\Omega_{2\gamma,\rm{in}}=0\,.
\ee
Note that the values for $\delta\Omega_{2\sigma}$ and
$\delta\Omega_{2\gamma}$ can exceed $1$, as can be seen in
Fig.~\ref{pix6}. As at first order, this is merely an artifact of
using normalised energy densities.
We do however see a new effect: at second order the energy densities
$\delta\rho_{2\alpha}$ can and do become negative, as can be seen
clearly in Figs.~\ref{pix5} and \ref{pix6}. This is a ``real'' effect
and not a normalisation artifact since $\rho_0\geq 0$ always. However,
the \emph{total} energy density $\rho$ as given by summing over all
the terms in the power series expansion \eq{rhosplit}, again stays
positive definite.

%%%%%%%%%%%%%%%%%%%%%%%%%%%%%%%%%%%%%%%%%%%%
\section{Relating the perturbative treatment 
to the $\Delta N$ formalism}
\label{deltaN_sect}
%%%%%%%%%%%%%%%%%%%%%%%%%%%%%%%%%%%%%%%%%%%%

The $\Delta N$ formalism \cite{SaSt95} provides a simple tool to
calculate the curvature perturbation on large scales at all orders in
the perturbations on scales larger than the horizon
\cite{SaTa,LMS,Lyth:2005fi}.
The main simplification compared to cosmological perturbation theory
stems from the fact that we only need the background evolution
equations, and not the full governing equations at all orders of
interest. However, if there is no analytic solution the numerics
necessary to get a result turn out to be quite involved as can be seen
below.
Nevertheless, we shall outline the calculation in the following.\\

The $\Delta N$ formalism relates the curvature perturbation on uniform
density hypersurfaces to the perturbation in the number of
e-foldings from the uniform density to the flat slicing,
\be
\label{defdeltaN}
\zeta=\delta N\,.
\ee

To get the number of e-foldings $N$ we use \eq{gdash} to get $d N$ in
terms of $d g$, and integrate,
\be
N = 2\int^{g_{\rm{fin}}}_{g_{\rm{in}}} 
\frac{dg}{g\(1-g\)\(4-\Omega_{0\sigma}\)} \,.
\ee
The curvature perturbation in the $\Delta N$ formalism is then given
from \eq{defdeltaN} by expanding $N$ in a Taylor series, which leads
in the curvaton case to \cite{Lyth:2005fi,Lyth2006}
\be
\label{deltaN2}
\zeta=
\frac{\partial N}{\partial \Omega_{0\sigma\rm{in}}} 
\delta\Omega_{\sigma\rm{in}}
+\frac{1}{2}
\frac{\partial^2 N}{\partial\Omega_{0\sigma\rm{in}}^2} 
\delta\Omega_{\sigma\rm{in}}^2  \,,
\ee
where the partial differentials are
\bea
\label{int1}
\frac{\partial N}{\partial \Omega_{0\sigma\rm{in}}} 
&=&2\int^g_{g_1} 
\frac{dg}{g\(1-g\)\(4-\Omega_{0\sigma}\)^2}
\frac{\partial\Omega_{0\sigma}}{\partial\Omega_{0\sigma\rm{in}}}  \,,\\
\label{int2}
\frac{\partial^2 N}{\partial\Omega_{0\sigma\rm{in}}^2} 
&=&
2\int^g_{g_{\rm{in}}} 
\frac{dg}{g\(1-g\)\(4-\Omega_{0\sigma}\)^2}\[
\frac{\partial^2\Omega_{0\sigma}}{\partial\Omega_{0\sigma\rm{in}}^2} 
+\frac{2}{\(4-\Omega_{0\sigma}\)}
\(\frac{\partial\Omega_{0\sigma}}{\partial\Omega_{0\sigma\rm{in}}}\)^2
\]  \,.
\eea
Note, that to make contact with first and second order perturbation
theory, $\zeta$ and $\delta\Omega_{\sigma\rm{in}}$ have to be expanded
up to second order, where in this case the second order curvature
perturbation corresponds to $\zetasb$, defined above in
\eq{defzetasb}, related to $\zeta_2$ as specified in \eq{defzeta2}.

Although in principle we can evaluate the integrals in \eqs{int1} and
(\ref{int2}) numerically and then differentiate them with respect to
the initial conditions to get the value of $\zeta$, this is (arguably)
more difficult than solving a set of coupled differential
equations. We therefore don't use the $\Delta N$ formalism in the
following sections, and solve instead the system of differential
equations presented in Section \ref{model_sect}.

%NEW v2
%
However, the $\Delta N$ formalism is used in
Ref.~\cite{misao_jussi_david} in another numerical study of the
curvaton scenario. The results are similar to the ones presented in
this paper, but the computing time required in the $\Delta N$ case is
increased by factor of roughly $\sim 100$ compared to solving the
system of differential equations presented in Section
\ref{model_sect}.

%%%%%%%%%%%%%%%%%%%%%%%%%%%%%%%%%%%%%%%%%%%%%%%%%%%%%%%%%%%%%%%%%%%%
\section{The non-linearity parameter $\fnl$: results and discussion}
\label{result_sect}
%%%%%%%%%%%%%%%%%%%%%%%%%%%%%%%%%%%%%%%%%%%%%%%%%%%%%%%%%%%%%%%%%%%

In this section we give the non-linearity parameter $\fnl$
calculated numerically using the governing equations at second order
of Sections \ref{gov_sect} and \ref{model_sect} and compare it to
previous numerical first order results and analytical sudden decay
estimates.\\

The non-gaussianity parameter $\fnl$ is defined as
\cite{Komatsu2001,Lyth:2005fi}
\be
\label{def_fnl}
\zeta=\zetag+\frac{3}{5}\fnl\(\zetag^2-{\bar{\zetag}}^2\)\,,
\ee
where $\zeta$ is the curvature perturbation at all orders, $\zetag$
the gaussian part of $\zeta$, and the ``bar'' denotes the spatial
average.
There has been some confusion in the literature as to the sign of
$\fnl$, which becomes relevant if the result is compared with
observations. The sign convention chosen here coincides with the one
used originally by Komatsu and Spergel \cite{Komatsu2001} and adopted
by most observational studies, and corrects the sign error introduced
in Ref.~\cite{LUW} and carried through in much subsequent work
\cite{Bartolo2003,Bartolo:2004if}
\footnote{Note that Eq.~(36) of Ref.~\cite{LUW}, corresponding to
\eqs{fnllinsd} and (\ref{fnllinnum}) here, has the correct sign for
$\fnl$, however there is a sign error in the derivation in
Ref.~\cite{LUW}.}.\\

We now relate the curvaton field fluctuations to the curvaton fluid
energy density.
The energy density in the curvaton field can be approximated by
\be
\label{rho_sigma}
\rho_\sigma=\frac{1}{2}m^2\sigma^2\,,
\ee
where $m$ is the curvaton mass and $\sigma$ is the amplitude of the
curvaton field.
Expanding the curvaton amplitude to first order, $\sigma=\sigma_0+
\delta\sigma_1$, we get from \eq{rho_sigma},
\bea
\label{defrhosigma}
&&\rho_{0\sigma}=\frac{1}{2}m^2\sigma_0^2\,,\\
\label{defdeltarhosigma1}
&&\frac{\delta\rho_{1\sigma}}{\rho_{0\sigma}}\equiv
2\frac{\delta\sigma_1}{\sigma_0}
+\(\frac{\delta\sigma_1}{\sigma_0}\)^2\,,\\
\label{defdeltarhosigma2}
&&\frac{\delta\rho_{2\sigma}}{\rho_{0\sigma}}\equiv 0\,.
\eea
Note, that including the quadratic term $(\delta\sigma_1/\sigma_0)^2$
in the first order energy density and setting the second order energy
density perturbation to zero is just convention, following
Ref.~\cite{LUW}. In Ref.~\cite{Bartolo2003} this term is included in
the second order energy density perturbation of the curvaton fluid.
This choice doesn't effect the final results.

%
%%%%%%%%%%%%%%%%%%%%%%%%%%%
\begin{figure}
\begin{center}
\includegraphics[angle=-0, width=88mm]{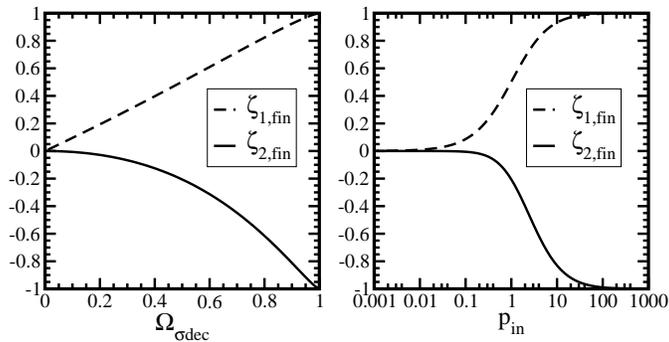} \\
\vspace{0.2cm}
\caption[pix]{
\label{pix12} 
The final values of the normalised curvature perturbations at first
and second order, $\zeta_{1,\rm{in}}$ and $\zeta_{2,\rm{in}}$, for an
initial value $\zeta_{1\sigma,\rm{in}}=1$, versus the background
curvaton energy density at decay, $\Omdec$, in the left panel and versus
$\pin$ in the right panel.}
\end{center}
\end{figure}
%%%%%%%%%%%%%%%%%%%%%%%%%%

%%%%%%%%%%%%%%%%%%%%%%%%%%
\subsection{Sudden decay}
%%%%%%%%%%%%%%%%%%%%%%%%%%

In the sudden decay one assumes that the curvaton doesn't decay into
radiation until a time $t_{\rm{dec}}$, when all of the curvaton energy
density decays suddenly; the normalised energy density of the curvaton
at decay is denoted $\Omdec$. The sudden decay approximation has been
widely used in the literature to study the curvaton scenario without
having to resort to numerical calculations, see
e.g.~Refs.\cite{curvaton,LUW,Bartolo2003,Lyth:2005fi}.
In order to be able to compare the sudden decay approximation to the
numerical calculation we now give a prescription to calculate
$\Omdec$.\\

The normalised background energy density can be approximated by
\cite{Linde2005,Lyth2006,curvaton}
\be
\label{omapprox}
\Omega_{0\sigma}
=\frac{\sigma_0^2}{\sigma_0^2+6\mpl\sqrt{\frac{H}{m}}}\,,
\ee
where $\mpl=(8\pi G)^{-1}$.
Assuming that the evolution of the background curvaton amplitude from
the initial time up to curvaton decay is negligible and using that
initially $H_{\rm{in}}=m$, and at curvaton decay $H_{\rm{dec}}=\Gamma$
we can use \eq{omapprox} to relate the parameter $\pin$, defined in
\eq{pin}, to the background energy density of the curvaton at
decay. We therefore define
\be
\label{def_omdec}
\Omdec\equiv\frac{\pin}{1+\pin}\,,
\ee
as energy density of the curvaton in the sudden decay approximation.
The agreement $\Omdec$ defined in \eq{def_omdec} with $\Omdec$ used in
Ref.~\cite{MWU} is quite good, and we use the definition
(\ref{def_omdec}) in the following to compare our numerical results
with the sudden decay approximation.\\

We now briefly review the results of previous analytical treatments
using the sudden decay approximation to calculate the non-linearity
parameter $\fnl$ in the curvaton scenario.

The non-linearity parameter in the sudden decay approximation using
first order perturbation theory, however using the definition of the
first order energy density perturbation quadratic in the curvaton
fluctuations \eq{defdeltarhosigma1}, is \cite{curvaton,LUW}
\be
\label{fnllinsd}
\fnl=\frac{5}{4\Omdec}\,.
\ee

Using second order perturbation theory the non-linearity parameter in
the sudden decay approximation was found to be
\cite{Bartolo2003,Lyth:2005du}
\be
\label{fnlsd2}
\fnl=\frac{5}{4\Omdec}-\frac{5}{6}\Omdec-\frac{5}{3}\,.
\ee

%%%%%%%%%%%%%%%%%%%%%%%%%%%%%%%%%
\subsection{Numerical solutions}
%%%%%%%%%%%%%%%%%%%%%%%%%%%%%%%%%

The transfer parameter at first order relating the initial curvature
perturbation on uniform curvaton density hypersurfaces to the final
value of the total curvature perturbation is defined as
\cite{curvaton,LUW,MWU}
\be
\label{def_r1}
r_1\equiv\frac{\zeta_{1,\rm{fin}}}{\zeta_{1\sigma,\rm{in}}}\,.
\ee
We define the transfer parameter at second order
\be
\label{def_r2}
r_2\equiv\frac{\zeta_{2,\rm{fin}} }{\zeta_{1\sigma,\rm{in}}^2}\,,
\ee
relating the final value of the total curvature perturbation to the
initial curvature perturbation on uniform curvaton slices.
The values for $r_1$ and $r_2$ coincide for our choice of initial
condition, $\zeta_{1\sigma,\rm{in}}=1$, with the final values of the
curvature perturbations at first and second order,
$\zeta_{1,\rm{fin}}$ and $\zeta_{2,\rm{fin}}$, and are given in
Fig.~\ref{pix12} versus $\Omdec$ and $\pin$.
% 
%%%%%%%%%%%%%%%%%%%%%%%%%%%
\begin{figure}
\begin{center}
\includegraphics[angle=0, width=100mm]{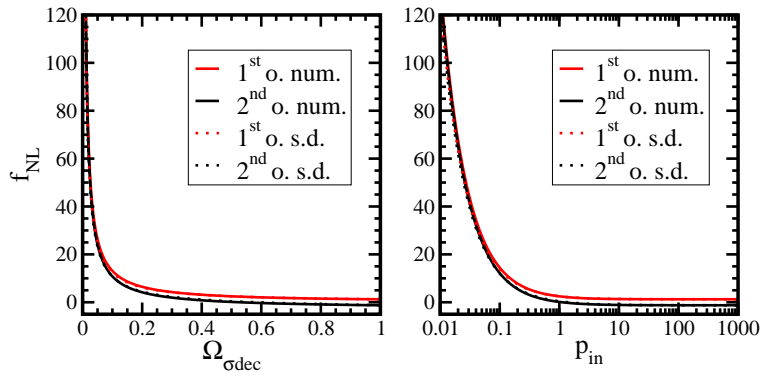} \\
\vspace{0.2cm}
\caption[pix]{
\label{pix8} 
The nonlinearity parameter $\fnl$ versus $\Omdec$ and $\pin$: numerical
results and sudden decay approximation at first and second order.}
\end{center}
\end{figure}
%%%%%%%%%%%%%%%%%%%%%%%%%%

The non-linearity parameter using first order perturbation theory is
given in terms of the transfer parameter defined in \eq{def_r1} as
\cite{LUW,MWU}
\be
\label{fnllinnum}
\fnl=\frac{5}{4 r_1}\,.
\ee

Using second order perturbation theory we find the non-linearity
parameter $\fnl$ from \eq{def_fnl}, expanding $\zeta$ to second order,
and get in terms of the transfer parameters at first and second order
\be
\label{fnl}
\fnl=\frac{5}{4r_1}+\frac{5}{6}\frac{r_2}{{r_1}^2}-\frac{5}{3}\,.
\ee
In the above calculations we identified $\zetag$ with the part of
$\zeta_1$ linear in the curvaton field fluctuation, i.e.~the first
term in \eq{defdeltarhosigma1}.\\
%
%%%%%%%%%%%%%%%%%%%%%%%%%%%
\begin{figure}
\begin{center}
\includegraphics[angle=0, width=100mm]{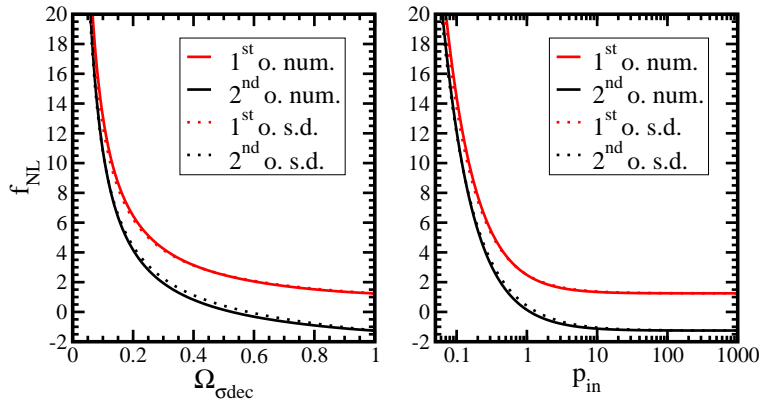} \\
\vspace{0.2cm}
\caption[pix]{
\label{pix9} 
The nonlinearity parameter $\fnl$ versus $\Omdec$ and $\pin$:
numerical results and sudden decay approximation at first and second
order, detail of Fig.~\ref{pix8}.
}
\end{center}
\end{figure}
%%%%%%%%%%%%%%%%%%%%%%%%%%

We can finally relate the transfer parameters $r_1$ and $r_2$ to the
total curvature perturbation, $\zeta=\zeta_1+\frac{1}{2}\zeta_2$,
evaluated after the curvaton has decayed,
\be
\zeta_{\rm{fin}}
=r_1 \zeta_{1\sigma,\rm{in}}
+\frac{1}{2} r_2\zeta_{1\sigma,\rm{in}}^2\,.
\ee
Using the definition of the curvature perturbation on uniform curvaton
density hypersurfaces, \eq{zeta1alpha}, and the expression for the
curvaton energy density in terms of the curvaton amplitude,
\eq{defdeltarhosigma1}, we get
\be
\zeta=\frac{2}{3}r_1\frac{\delta\sigma_1}{\sigma_0}
+\frac{1}{3}\(r_1+\frac{2}{3}r_2\)
\(\frac{\delta\sigma_1}{\sigma_0}\)^2\,.
\ee
We can compare these results to the expression found using the $\Delta
N$-formalism, expressed in terms of the curvaton perturbation (instead
of the normalised energy perturbation used in Section
\ref{deltaN_sect}) \cite{Lyth:2005fi,Lyth2006},
\be
\zeta_{\rm{SB}}=N_{,\sigma}\delta\sigma
+\frac{1}{2}N_{,\sigma\sigma}\delta\sigma^2\,.
\ee
The curvature perturbation employed in the $\Delta N$-formalism,
$\zetasb$, is related to $\zeta$ by \eq{defzetasb}, and we get
\be
\zeta_{\rm{SB}}=\frac{2}{3}r_1\frac{\delta\sigma_1}{\sigma_0}
+\frac{1}{3}\(r_1+\frac{2}{3}r_2-\frac{4}{3}r_1^2\)
\(\frac{\delta\sigma_1}{\sigma_0}\)^2\,,
\ee
and therefore 
\be
N_{,\sigma} = \frac{2}{3}\frac{r_1}{\sigma_0}\,,
\qquad N_{,\sigma\sigma}
=\frac{2}{3\sigma_0^2}\(r_1+\frac{2}{3}r_2-\frac{4}{3}r_1^2\)\,.
\ee

%%%%%%%%%%%%%%%%%%%%%%%%%%%%%%%%%%%%%%%%%
\subsection{Curvaton amplitude evolution}
\label{curvaton_evol}
%%%%%%%%%%%%%%%%%%%%%%%%%%%%%%%%%%%%%%%%%

So far we assumed that the curvaton field doesn't evolve between the
end of inflation and curvaton decay.
In order to allow for the evolution of curvaton field amplitude
$\sigma$ we assume that it depends on the initial value set during
inflation $\se$ by $\sigma=\sigma(\se)$ which gives for the curvaton
field fluctuation \cite{Lyth:2003dt,Enqvist:2005pg,Lyth2006}
\be
\delta\sigma=\sigma'\delta\se+\frac{1}{2}\sigma''\delta\se^2\,,
\ee
where $\sigma'\equiv\p\sigma/\p\se$. For the first order energy
density we then find (including again the quadratic term)
\be
\frac{\delta\rho_{1\sigma}}{\rho_{0\sigma}}=
2\frac{\delta\sigma_{\rm{e}1}}{\sigma_0}\sigma_{0}'
+\(\frac{\delta\sigma_{\rm{e}1}}{\sigma_0}\)^2
\(1+\sigma_0\frac{\sigma_{0}''}{{{\sigma_{0}'}^2}}\){\sigma_{0}'}^2\,.
\ee
We therefore get for the non-linearity parameter 
\be
\label{fnl_evol}
\fnl=\[\frac{5}{4r_1}+\frac{5}{6}\frac{r_2}{{r_1}^2}\]
\(1+\sigma_0\frac{\sigma''_{0}}{{{\sigma'_{0}}^2}}\)
-\frac{5}{3}\,,
\ee
instead of \eq{fnl} above. However, in order to calculate a numerical
value for $\fnl$ we now have to calculate the evolution of $\sigma$ in
detail and specify a curvaton model. We shall therefore not pursue
this issue further and refer to Ref.~\cite{Enqvist:2005pg} where this
issue was studied in detail.

%%%%%%%%%%%%%%%%%%%%%%%%%%%%%%%%%%%
\subsection{Results and discussion}
%%%%%%%%%%%%%%%%%%%%%%%%%%%%%%%%%%%

\begin{figure}
\begin{center}
\includegraphics[angle=0, width=100mm]{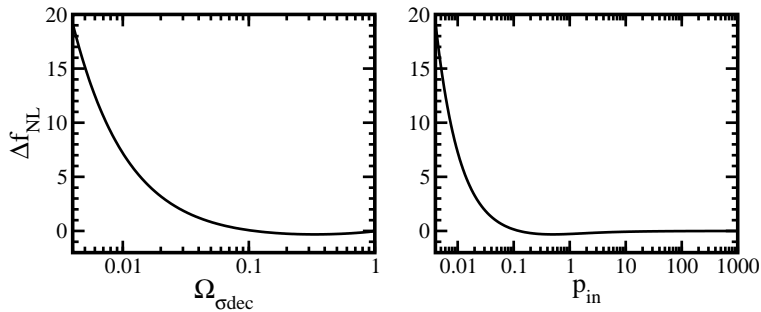} \\
\vspace{0.2cm}
\caption[pix]{
\label{pix10} 
The difference of the numerical and the sudden decay approximation
value of the non-linearity parameter, $\Delta\fnl$, versus $\Omdec$
and $\pin$.}
\end{center}
\end{figure}

Our results are summed up in Figs.~\ref{pix8}-\ref{pix10}:
in Figs.~\ref{pix8} and \ref{pix9} we plot the non-linearity
parameter calculated numerically using \eq{fnllinnum} at first order
and \eq{fnl} at second order for two different parametrisations,
namely $\Omdec$and $\pin$.
In the same figures we also plot the non-linearity parameter $\fnl$ in
the sudden decay approximation at first and second order from
\eqs{fnllinsd} and (\ref{fnlsd2}), respectively. We truncated the
graphs at $\fnl=120$ in accordance with the current observational
bounds (see below).
We first note how well the sudden decay approximation and the
numerical solution at first and second order agree. However, at first
order we find that $\fnl=5/4$ for $\Omdec = 1$ or large values of
$\pin$, whereas including the second order effects we get $\fnl=-5/4$
for $\Omdec = 1$ or large values of $\pin$.

In Fig.~\ref{pix10} we plot the difference between the non-linearity
parameter obtained numerically and using the sudden decay
approximation, both at second order,
\be
\Delta\fnl\equiv
\fnl\Big|_{\rm{numerical}} - \fnl\Big|_{\rm{sudden decay}}\,.
\ee
We now see more clearly the excellent agreement of the sudden decay
approximation for large parameters $\pin$ and $\Omdec$. However, for
small $\pin$ and $\Omdec$ the sudden decay approximation works less
well, deviating from the numerical solution by up to $10\%$.

%%%%%%%%%%%%%%%%%%%%%%%%%%%
%%%%%%%%%%%%%%%%%%%%%%%%%%
%
We finally give the observational constraint on the non-linearity
parameter $\fnl$ from the recently published WMAP three-year data.
Spergel et al.~found \cite{Spergel}: $-54<\fnl<114$ (at $95\%$
confidence level). The curvaton model is therefore well within the
current observational bounds.
However, if future observations give a large negative value for the
non-linearity parameter, the curvaton model would be ruled out, at
least without strong evolution of the curvaton amplitude from the
beginning of the oscillations to curvaton decay, as pointed in in
Section \ref{curvaton_evol} above.

%%%%%%%%%%%%%%%%
\acknowledgments
%%%%%%%%%%%%%%%%

The authors are grateful to Jussi Valiviita and David Wands for useful
comments. KAM is supported by PPARC grant PPA/G/S/2002/00098, DHL is
supported by PPARC grants PPA/V/S/2003/00104, PPA/G/O/2002/00098 and
PPA/S/2002/00272 and EU grant MRTN-CT-2004-503369.
Algebraic computations of tensor components were performed using
the \textsc{GRTensorII} package for Maple.

%%%%%%%%%%%%%%%%%%%%%%%%%%%%%%%%%%%%%%%%%%%%%%%%%%%%%%%%%%%%%%%%%%%%%%%%

%%%%%%%%%%%%%%%%%%%%%%%%%%%%%%%%%%%%%%%%%%%%%%%%%%%%%%%%%%%%%%%%%%%%%%%%

\appendix
%%%%%%%%%%%%%%%%%%%%%%%
\section{Governing equations}
%%%%%%%%%%%%%%%%%%%%%%%

Here we first give the governing equations on large scales in the
general case without gauge restrictions and then the equations given
in Section \ref{model_sect} in terms of non-normalised energy
densities.

%%%%%%%%%%%%%%%%%%%%%%%%%%%%%%%%
\subsection{Governing equations without gauge restriction}
%%%%%%%%%%%%%%%%%%%%%%%%%%%%%%%%

In this subsection we give the governing equations on large scales in
the general case without any gauge restrictions, i.e.~without choosing
a particular hypersurface.

%%%%%%%%%%%%%%%%%%%%%%%%%%
\subsubsection{First order}
\label{governing1}
%%%%%%%%%%%%%%%%%%%%%%%%%%

Energy conservation of the $\alpha$-fluid is given from
\eq{nablaTalpha} at first order as
\be 
\label{dotrhoalphanogauge1}
\dot{{\delta\rho_{1\alpha}}}
+3H\left( {\delta\rho_{1\alpha}}+{\delta P_{1\alpha}}\right)
-3\dot\psi_1\left( \rho_{0\alpha}+P_{0\alpha}\right)
-Q_{0\alpha}\phi_1 - {\delta Q_{1\alpha}}=0\,.
\ee
Total energy conservation follows from \eq{dotrhoalphanogauge1} above, and
using \eqs{Qconstraint} and (\ref{defrhotot}), is given by
\be 
\dot{{\delta\rho_{1}}}
+3H\left( {\delta\rho_{1}}+{\delta P_{1}}\right)
-3\dot\psi_1\left( \rho_{0}+P_{0}\right)
=0\,.
\ee
The $0-0$ Einstein equation is given from \eq{Einstein} as
\be
3H\left( H\phi_1+ \dot\psi_1\right) = -4\pi G \delta\rho_1\,.
\ee

%%%%%%%%%%%%%%%%%%%%%%%%%%
\subsubsection{Second order}
\label{governing2}
%%%%%%%%%%%%%%%%%%%%%%%%%%

Energy conservation of the $\alpha$-fluid is given from
\eq{nablaTalpha} at second order as
\bea 
\dot{{\delta\rho_{2\alpha}}}
&+&3H\left( {\delta\rho_{2\alpha}}+{\delta P_{2\alpha}}\right)
-3\dot\psi_2\left( \rho_{0\alpha}+P_{0\alpha}\right)
-6\left(\delta\rho_{1\alpha}+\delta P_{1\alpha} \right)\dot\psi_1
-12\left( \rho_{0\alpha}+P_{0\alpha}\right)\dot\psi_1\psi_1\nonumber\\
&-&Q_{0\alpha}\left(\phi_2-\phi_1^2\right) 
-2\phi_1{\delta Q_{1\alpha}} - {\delta Q_{2\alpha}}
=0\,,
\eea
and, following a similar route as at first order, the conservation of
the total energy density is given at second order by
\bea 
\dot{{\delta\rho_{2}}}
&+&3H\left( {\delta\rho_{2}}+{\delta P_{2}}\right)
-3\dot\psi_2\left( \rho_{0}+P_{0}\right)
-6\left(\delta\rho_{1}+\delta P_{1} \right)\dot\psi_1
-12\left( \rho_{0}+P_{0}\right)\dot\psi_1\psi_1
=0\,,
\eea
and the $0-0$ Einstein equation is given by
\be
3H^2\left(\phi_2-4\phi_1^2\right)+3\left(H\dot\psi_2-{\dot\psi_1}^2\right)
+12H\dot\psi_1\left(\psi_1-\phi_1\right)=-4\pi G \delta\rho_2 \,.
\ee

%%%%%%%%%%%%%%%%%%%%%%%%%%%%%%%%%%%%%%%%%%%%%%%%%%%%%%%%%%%%%%%%%%%%%%%%%
\subsection{Governing equations in terms of non-normalised energy densities}
\label{gov_equ_non_non}
%%%%%%%%%%%%%%%%%%%%%%%%%%%%%%%%%%%%%%%%%%%%%%%%%%%%%%%%%%%%%%%%%%%%%%%%%

In this subsection we give the governing equations presented in
Sections \ref{gov_equ1} and \ref{gov_equ2} above in terms of the
normalised quantities in terms of the non-normalised energy densities
and decay rate. We use the number of e-foldings $N$ as time coordinate
and work throughout in the flat gauge (omitting the ``tilde'').\\

%%%%%%%%%%%%%%%%%%%%%%%%%%%%
%\subsubsection{First order}
%%%%%%%%%%%%%%%%%%%%%%%%%%%%

We get at first order
\bea
&&\delta\rho_{1\sigma}'+\left(3+\frac{\Gamma}{H}\right)\delta\rho_{1\sigma}
-\frac{1}{2}\frac{\Gamma}{H}\frac{\rho_{0\sigma}}{\rho_0}\delta\rho_1=0\,,\\
&&\delta\rho_{1\gamma}'+4\delta\rho_{1\gamma}
-\frac{\Gamma}{H}\delta\rho_{1\sigma}
+\frac{1}{2}\frac{\Gamma}{H}\frac{\rho_{0\sigma}}{\rho_0}\delta\rho_1=0\,,
\eea
%
%%%%%%%%%%%%%%%%%%%%%%%%%%%%
%\subsubsection{Second order}
%%%%%%%%%%%%%%%%%%%%%%%%%%%%
%
and at second order
\bea
&&\delta\rho_{2\sigma}'+\left(3+\frac{\Gamma}{H}\right)\delta\rho_{2\sigma}
-\frac{\Gamma}{H}\frac{\delta\rho_1}{\rho_0}\delta\rho_{1\sigma}
+\frac{\Gamma}{H}
\frac{\rho_{0\sigma}}{\rho_0}
\left(
\frac{3}{4}\frac{{\delta\rho_1}^2}{\rho_0}
-\frac{1}{2}\delta\rho_2\right)
=0\,,\\
&&\delta\rho_{2\gamma}'+4\delta\rho_{2\gamma}
-\frac{\Gamma}{H}\delta\rho_{2\sigma}
+\frac{\Gamma}{H}\frac{\delta\rho_1}{\rho_0}\delta\rho_{1\sigma}
-\frac{\Gamma}{H}
\frac{\rho_{0\sigma}}{\rho_0}
\left(
\frac{3}{4}\frac{{\delta\rho_1}^2}{\rho_0}
-\frac{1}{2}\delta\rho_2\right)
=0\,.
\eea

%%%%%%%%%%%%%%%%%%%%%%%%%%%%
{}
%%%%%%%%%%%%%%%%%%%%%%%%%%

%%%%%%%%%%%%%%%%%%%%%%%%%%%%%%%
\end{document}